\documentclass[reprint,unsortedaddress,amsmath,amssymb,aps,prl,showpacs]{revtex4-2}
\usepackage{amsmath}%
\usepackage{amsfonts}%
\usepackage{amssymb}%
\usepackage[cal=cm]{mathalfa}
\usepackage{graphicx}
\usepackage{dcolumn}
\usepackage{bm}
\usepackage{braket}
\usepackage{sidecap}
\usepackage{color}
\usepackage{bbm}
\usepackage{nicefrac}
\usepackage{float}
\usepackage{placeins}
\usepackage[dvipsnames]{xcolor}
\usepackage[utf8]{inputenc}
\usepackage[normalem]{ulem}
\usepackage{nicefrac}
\usepackage{dsfont}
\usepackage{xcolor, soul} 
\usepackage[normalem]{ulem} 
\usepackage{siunitx}
\usepackage[citebordercolor=blue]{hyperref}
\usepackage[utf8]{inputenc}
\usepackage{gensymb}
\usepackage{mathtools}

\begin{document}

\title{CANISIUS The Austrian Neutron Spin Echo Interferometer}

\author{Niels Geerits$^{1}$}
\email{niels.geerits@tuwien.ac.at}
\author{Simon Hack$^1$}
\author{Lara Brukner$^{1}$}
\author{Ad van Well$^2$}
\author{Steven R. Parnell$^{2,3}$}
\author{Hartmut Abele$^1$}
\author{Stephan Sponar$^1$}
\email{stephan.sponar@tuwien.ac.at}
\affiliation{%
$^1$Atominstitut, Technische Universit\"at Wien, Stadionallee 2, 1020 Vienna, Austria \\
$^2$Faculty of Applied Sciences, Delft University of Technology, Mekelweg 15, Delft 2629JB, The Netherlands \\
$^3$ISIS, Rutherford Appleton Laboratory, Chilton, Oxfordshire, OX11 0QX, UK}

\date{\today}
\hyphenpenalty=800\relax
\exhyphenpenalty=800\relax
\sloppy
\setlength{\parindent}{0pt}
\begin{abstract} 
The broad-band resonant spin echo interferometer, CANISIUS, is presented. CANISIUS is built in a versatile way, such that it can be operated in both a continuous broad-band beam or a pulsed Time of Flight beam. This versatility also extends to the modes available to the instrument, such as Neutron Resonant Spin Echo, Spin Echo (Modulated) Small Angle Neutron Scattering and coherent averaging to produce structured wavefunctions for scattering. The instrument may also be used as an interferometer, to probe fundamental questions in quantum mechanics. In this paper we detail both the continuous and Time of Flight options of the instrument. In addition we demonstrate the applicability of our interferometer to ultra small angle scattering in a white beam. Finally we demonstrate a new spin echo interferometry tool, which uses incomplete recombination of the two path states to generate composite wavefunctions with special structure. In particular we show that this method produces neutron wavefunctions that exist in a superposition of two quantum mechanical orbital angular momentum modes, $\ell=\pm 1$. We illustrate that just as this method can be used to generate certain structured waves, it may also be used to characterize the structure of the input wavefunction.
\end{abstract}

\maketitle
\section{Introduction}
Neutron interferometry is a powerful and established tool for addressing a variety of scientific challenges from exploring quantum state entanglement in massive particles \cite{Hasegawa2003,Hasegawa2010} and inertial effects on quantum particles such as gravitationaly induced interference (COW experiment) \cite{Colella1975}, the Sagnac effect \cite{Sagnac1913,Werner1979} and the Mashhoon effect \cite{Mashhoon1988,Danner2020}, to nuclear physics questions by measuring scattering cross sections to high precision \cite{Haun2020}. Furthermore interferometers may be used to create complex structured waves by coherent averaging \cite{Sarenac2018b,Geerits2023}. This method is particularly useful for producing vortex neutrons \cite{Sarenac2019,Geerits2023}, which carry Orbital Angular Momentum (OAM) and may have unique scattering properties \cite{Afanasev2019,Afanasev2021,Jach2022}. Furthermore, OAM represents an additional quantum mechanical degree of freedom, which is useful for quantum information and contextuality experiments.
\par 
The most well known type of neutron interferometer is made from a perfect silicon crystal in triple Laue configuration \cite{Rauch1974}, however other methods have been established as well such as supermirror based approaches \cite{Fujiie2024} and magnetic refraction based approaches \cite{Gaehler1996}. Neutron spin echo \cite{Mezei1972} techniques which use the latter method have been particularly successful. The classic spin echo technique, resonant spin echo \cite{Golub1987} and "Modulation of Intensity Emerging from Zero Effort" (MIEZE) \cite{Gahler1992} interfere the two neutron spin states in time, which makes these techniques particularly sensitive to dynamic effects, such as those observed in inelastic scattering. In addition MIEZE has even been applied to observe violations of Bell's inequality \cite{Sponar2010,Leiner2024}. In 1978 Pynn proposed a technique to apply spin echo to ultra small angle scattering, by inclining the magnetic field regions with respect to the direct beam \cite{Pynn1978}. Almost two decades later, Rekveldt proposed to utilize shaped poleshoes in addition to foil flippers to increase the usable field integral. Over time the technique became known as spin echo small angle neutron scattering (SESANS). This method splits and later focuses the two spin states transversely, which makes the instrument sensitive to structure along the refraction direction (defined by the inclination of the magnetic field region) \cite{Gaehler1996}. In recent years many of the famous neutron interferometry experiments, such as the COW experiment \cite{Haan2014,Parnell2020} and tests of quantum entanglement \cite{Shen2020,Kuhn2021} have been carried out with spin-echo type interferometers. In addition these interferometers have a few advantages over perfect crystal interferometers, such as the ability to accept a large, broad-band and wide divergence beam. As a result spin-echo interferometer have a higher usable flux. In addition, this class of interferometer is not as sensitive to temperature variations and vibrations. The most obvious disadvantage is the microscopic path separation (10 $nm$ to 10 $\mu m$), consequently the paths can usually not be manipulated separately. Nonetheless, the area spanned by the two paths of a spin-echo interferometer is similar to that of a single crystal interferometer. Also, it is simpler to produce structured wavefunctions by coherent averaging with spin echo interferometers, due to the possibility of a modular design, which allows for an arbitrary number of beam splitters and paths in a single instrument. \par
\begin{figure*}
	\includegraphics[width=18cm]{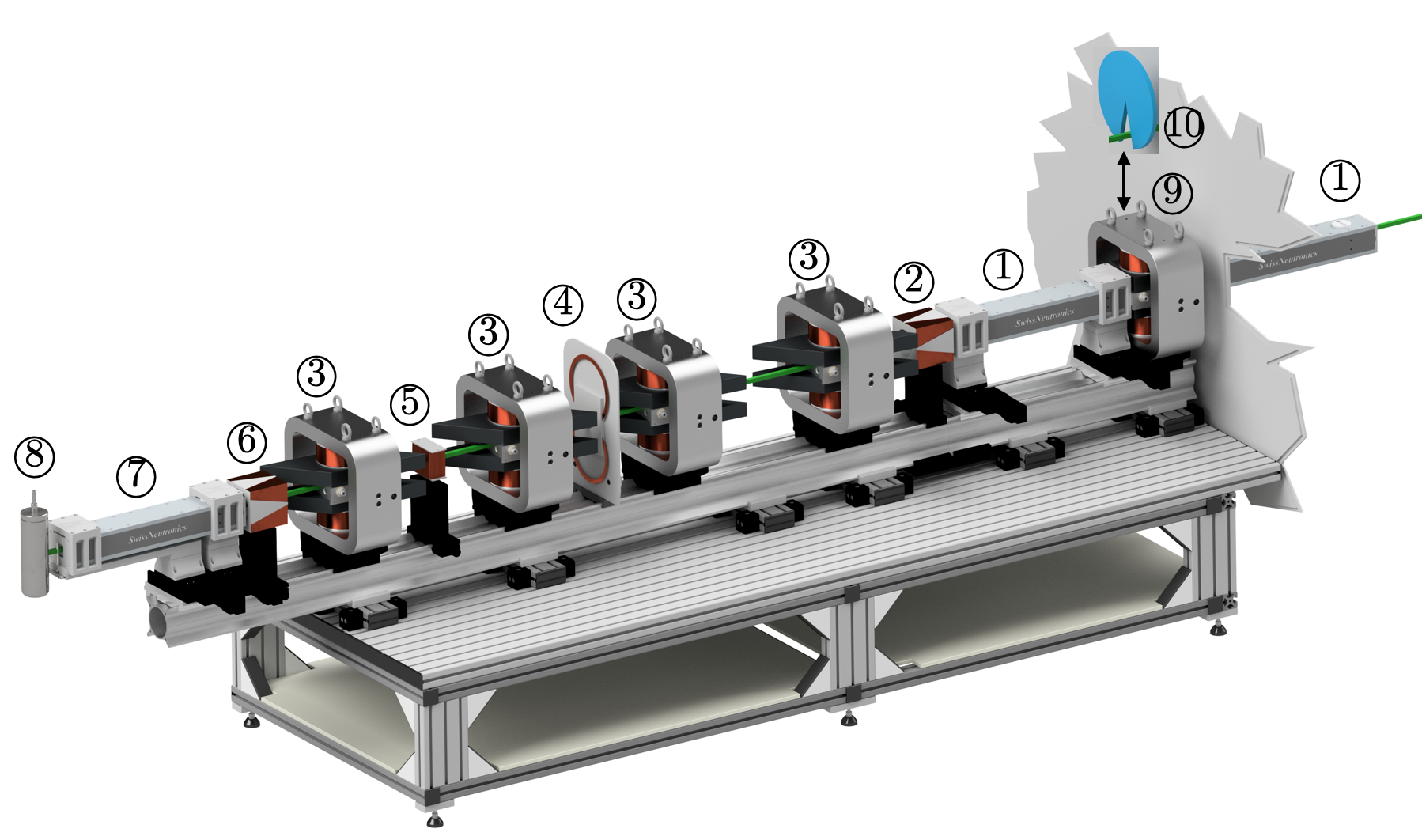}
	\caption{Render of the CANISIUS instrument configured for white beam SESANS. The beam propagates from right to left. First the beam is polarised by double reflection from two single substrate m=4 polarising supermirrors (1). Next the polarisation is adiabatically rotated by 90 degrees by a v-coil (2). The beam then passes through a pair (arm 1) of adiabatic RF flippers with parallelogram shaped poleshoes, which act as the first beam splitter and mirror (3). A field stepper is positioned between the first and second pair (arm 2) of RF flippers, to facilitate a fast non-adiabatic field transition from arm 1 to arm 2 (4). \textcolor{black}{Samples of various types may be inserted right before the field stepper.} The second arm of RF flippers serve effectively as a mirror and beam splitter to recombine the split beams. Between the last two RF flippers the beam is passed through a precession coil (5), which manipulates the spin phase to measure the spin echo curve. Finally before the spin is selected by the last polarising supermirror (7) the neutrons pass through a final v-coil facilitating another adiabatic 90 degree rotation (6). \textcolor{black}{Neutrons passing the  polarising supermirror are detected by a high-efficency $^3\textrm{He}$ counting tube. In addition to the usual continuous mode of operation the CANISIUS instrument can also be operated in ToF mode. For this purpose two chopper device are available: a conventional mechanical chopper (10) as well as a spin-chopper system (9), consisting of an RF flipper in combination with the pre-polarizer, that is the very first supermirrors (1). The length of the instrument, given by the distance between chopper (9,10) and detector (8) is 3 meters.} }\label{Setup}
\end{figure*}
In this paper we report on the new broad-band spin-echo interferometer, CANISIUS (Coherent Averaging Neutron Instrument for Spin-echo Interferometry and fUndamental Science), designed to investigate the properties of vortex neutrons carrying non-zero OAM. CANISIUS follows similar design doctrines as the Offspec \cite{Dalgliesh2011} and Larmor \cite{Geerits2019,Geerits2019b} instruments at the ISIS neutron source, as well as the instruments developed at TU Delft \cite{Rekveldt2005}. That is to say, CANISIUS uses adiabatic RF flippers \cite{Bazhenov1993} to ensure broad-band efficiency and is built in a modular way so that it can be applied to a variety of different modes (SESANS, SEMSANS \cite{Li2016}, NRSE and MIEZE). In addition CANISIUS can quickly change between a continuous white beam and a pulsed beam for time of flight. Here we will report on the design and efficiency of the adiabatic RF-flippers and the instrument, the interference patterns produced by the instrument in broad-band and time of flight (ToF) SESANS modes and finally discuss production and investigation of vortex neutrons.
\section{Methods}
\subsection{Instrument Overview}
A three dimensional render of the CANISIUS instrument in a SESANS configuration is shown in figure \ref{Setup}. CANISIUS is situated at the white beamline of the Atominstitut (TU Wien) 250 kW reactor. The instrument can be operated in either a continuous broad-band or two distinct ToF modes. A standard mechanical chopper is used for the regular ToF mode. The second mode is enabled by a spin chopping system, which is made possible since the beam is initially polarized by double reflection from a pair of m=4 single flat substrate supermirrors, this allows the beam to be chopped by inserting a broad-band spin flipper between the two mirrors which can be pulsed \cite{Rauch1968,Rauch1972}. Due to the cut-off of the supermirrors the available wavelength range is $2 \mathrm{\AA}-6 \mathrm{\AA}$, with a peak at $2.5 \mathrm{\AA}$. After the double reflection polarizer the beam is passed through a v-coil which adiabatically rotates the neutron spin around the propagation axis from vertical to horizontal. In the first arm of the instrument which consists of a pair of adiabatic RF flippers \cite{Grigoriev2001}, the two spin states are separated longitudinally (standard NRSE) and if parallelogram shaped poleshoes are employed also transversely (SESANS). The second arm with equal but opposite field of the first arm recombines the spin states. Spatial overlap between the two spin states is ultimately measured by a spin projective measurement, which in practice is enabled by a v-coil and a supermirror. An additional closed coil in arm 2, allows variation of the phase between the two spin states. The polarisers and RF coils define the maximal beam size to 7 x 20 mm$^2$. This beam size is used for most measurements described here, except when otherwise specified. A calibration wedge can be inserted between the yoke and the poleshoe to change and calibrate the angle of the poleshoes by up to $\pm 5$ degrees. This gives some more flexibility to tune the transverse spin state separation, however its primary purpose is to enable the new coherent averaging mode, to produce structured neutron waves. If both arms of the interferometer are operated at a different poleshoe angle the instrument fails to focus the spin states transversely. As a result some residual separation is left between the spin states. As we will show in a later section, this residual separation induces a structured wave with OAM. The residual separation may also be scanned, thus allowing one to measure the correlation between the spin up and down wavefunctions, similar to what is described in \cite{Rauch1996}. On average over the entire wavelength band the polarisation of the instrument in the non-echo mode (RF system and v-coils turned off), in which there should be no spin precession, is 0.904. The RF coils are matched to the output impedance of the amplifiers for operation at 1.4 MHz, however the coils may also be operated in resonant mode (ToF or monochromatic), in which case they can be operated at any frequency between 10 kHz and 5 MHz, since the power requirements are low (1-2 W) and the amplifiers are equipped to handle up to 400 W of reflected power. In the following subsections the adiabatic RF flippers, SESANS, the ToF modes and the coherent averaging mode will be discussed in more detail.
\subsection{Adiabatic RF Spin Flippers}
\begin{figure}[!t]
 \includegraphics[width=8cm]{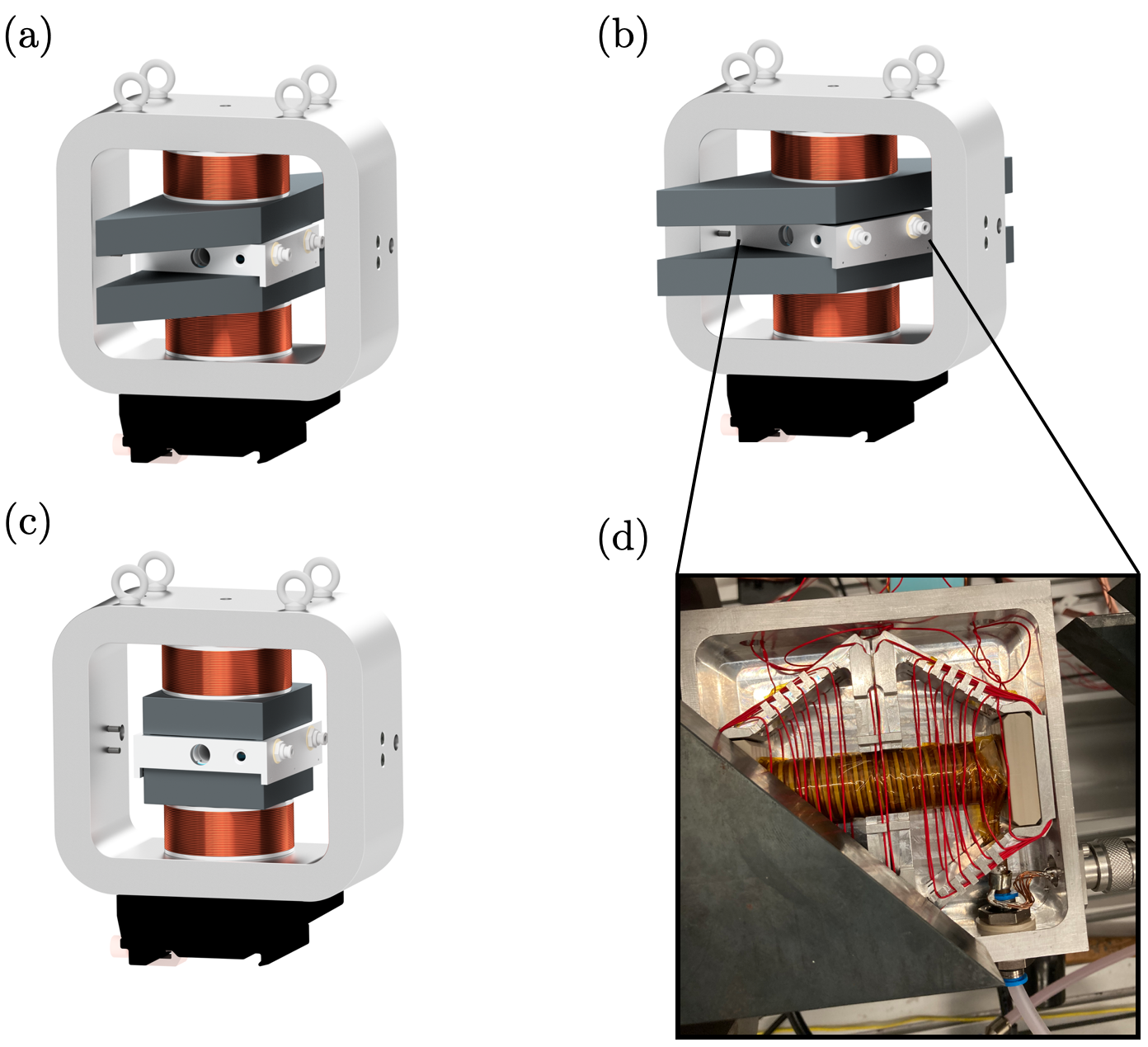}
	\caption{\textcolor{black}{Render of the adiabatic RF spin flippers with parallelogram shaped poleshoes at 45\,deg (a), 45\,$\pm$ 5\,deg (b), and 90 deg in (c). The photo in (d) shows a top-view of one  RF coil partly taken out of the parallelogram shaped poleshoes for demonstration purposes to see the water cooled RF coil, the gradient field coil, as well as the inlet of the coolant and the electrical conducting.    }  }\label{RFs}
\end{figure}
\begin{figure}
 \includegraphics[width=9cm]{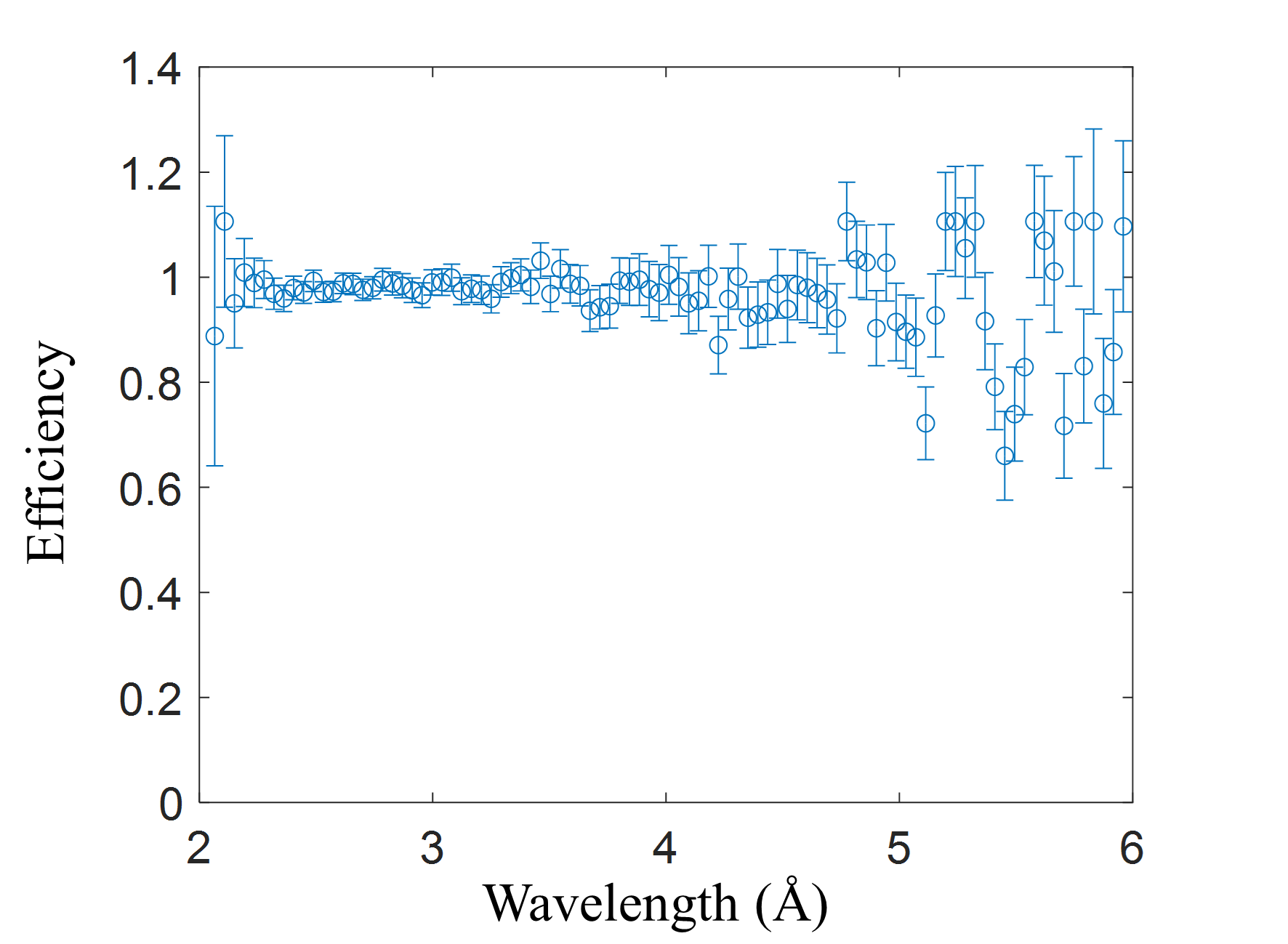}
	\caption{Flipping efficiency of one adiabatic RF flipper against wavelength measured using the ToF technique. This is calculated using ToF spectrums with the flipper on and off, the result is finally normalised by the instrument efficiency of 0.904. The weighted average of the flipping effiency over the entire spectrum is 0.977.}\label{efficiency}
\end{figure}

In this subsection we provide a detailed overview of the design of CANISIUS' RF flipper system. An RF flipper is depicted in figure \ref{RFs}, for 3 different poleshoe configurations, square for NRSE and MIEZE, parallelogram for SESANS and SEMSANS and tilted parallelgram for coherent averaging. The poleshoes in addition to the magnetic cores were machined from soft iron and subsequently heat treated. The yokes were machined from construction steel. Coils of 400 windings each are situated around each magnetic core. The gap between the poleshoes is 30 mm, in which the RF system is situated (figure \ref{RFs} (d)). This RF system consists of an aluminium housing containing an 80 mm long RF coil with a diameter of 20 mm and a gradient coil which has minimal winding density in the center and increases towards in the ends of the coil. The RF coil consists of 2 mm copper tubing (inner diameter 1.5 mm) wound on a PEEK bobbin and encased in kapton foil to prevent arcing. 250 ml/min of G12 coolant is pumped through the tubing, to facilitate heat transport, the coolant is cooled to 3 degrees by a secondary loop connected to a chiller. Matching of the RF coils to the amplifier output is done capacitvely for a single frequency, 1.4 MHz. Mica capacitors were used for their high stability and low loss at typical NRSE frequencies. In adiabatic flipping mode, roughly 60 W are dissipated in each coil, in addition to 40 W in each gradient coil. In resonant flipping mode only 1-2 W are required depending on the frequency, hence in this mode matching between load and amplifier is not necessary. The efficiency of a single adiabatic RF flipper is shown in figure \ref{efficiency}, which is determined first by measuring the polarisation in a non precessing mode (i.e. v-coil turned off) and then normalizing said polarization by the the instrument spin transport efficiency, which as mentioned earlier is 0.904. The polarization is calculated using the well known formula
\begin{equation}
    P=\frac{I_+-I_-}{I_++I_-}
\end{equation}
with $I_+$ the intensity of the spin state aligned to the analyzing direction (i.e. when the flipper is turned on) and $I_-$ the opposite spin state.
Figure \ref{efficiency} demonstrates that our RF flippers have a constant flip efficiency over the entire wavelength band indicating a good adiabicity of the flippers. A weighted average efficiency of 0.977 can be extracted from the data shown in figure \ref{efficiency}. This is comparable to some other adiabatic RF flipper designs \cite{Li2020}. Though we note that flipping efficiency of $~99\%$ are possible with similar design \cite{Plomp2009}. In some other desings efficiencies of up to $99.9\%$ have been reported \cite{Chen2021}.
\subsection{Time of Flight Options}
\begin{figure}
 \includegraphics[width=9cm]{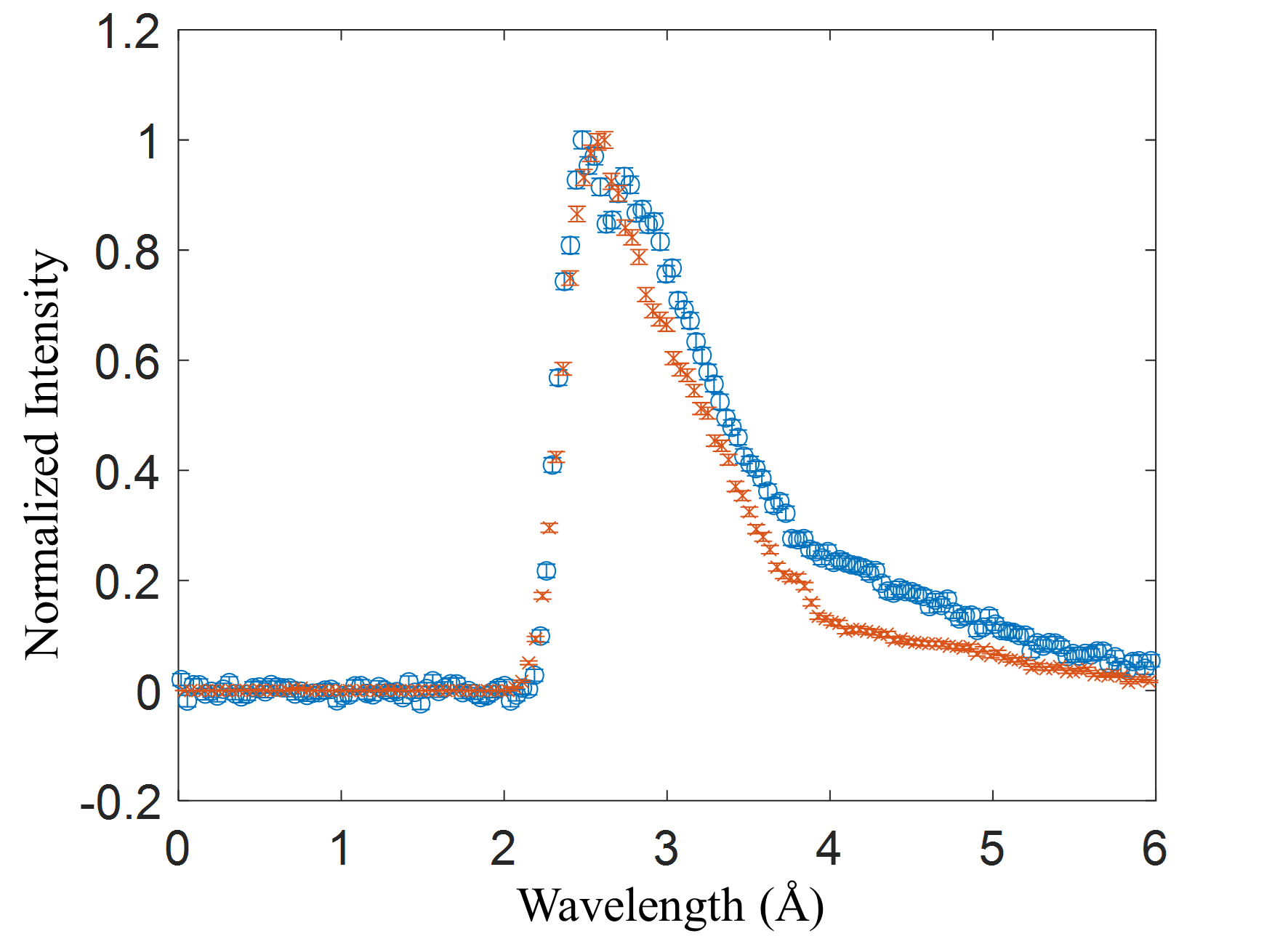}
	\caption{Normalized Time of Flight spectra produced using the RF spin chopping device (blue circular data points) and the mechanical chopper (red cross shaped data points).}\label{TOF}
\end{figure}

CANISIUS has two options available for producing a pulsed beam. The first is a Fermi like chopper which consists of stacked, short, straight, cadmium lined channels, which is spun at 50 Hz, resulting in an effective pulse frequency of 100 Hz. The second method, uses an RF flipper positioned between the two polarizing supermirrors. In this case the the two mirrors are oriented to select opposite spins, so that when the flipper is off no beam is transmitted into the instrument. In resonant mode the phase accumulated by a neutron in the flipper is given by
\begin{equation}
    \alpha=\gamma B_{RF} t
\end{equation}
with $\gamma$ the gyromagnetic moment of the neutron $B_{RF}$ the field strength and $t$ the time for which the particle is exposed to the field. Hence if the RF pulse time is much shorter than the flight time through the device of the fastest neutron then this phase is wavelength independent. Therefore to produce a pulsed beam the RF flipper is simply shortly pulsed in resonant mode, which flips all neutrons that are in the RF coil at the time of the pulse. These neutrons can then be transmitted through the instrument and to the detector. This RF flipper is designed equivalently to the others described in the previous section, with the exception of the gradient coil, since this is not required. The RF field pulse length is 5 $\mu s$, ten times shorter than the flight time of $2.5 \mathrm{\AA}$ neutrons through the RF coil. This corresponds to 7 cycles of the RF field at 1.4 MHz. In figure \ref{TOF} a comparison between the ToF spectra of the mechanical and spin chopper can be seen.
\begin{figure*}
	\includegraphics[width=18cm]{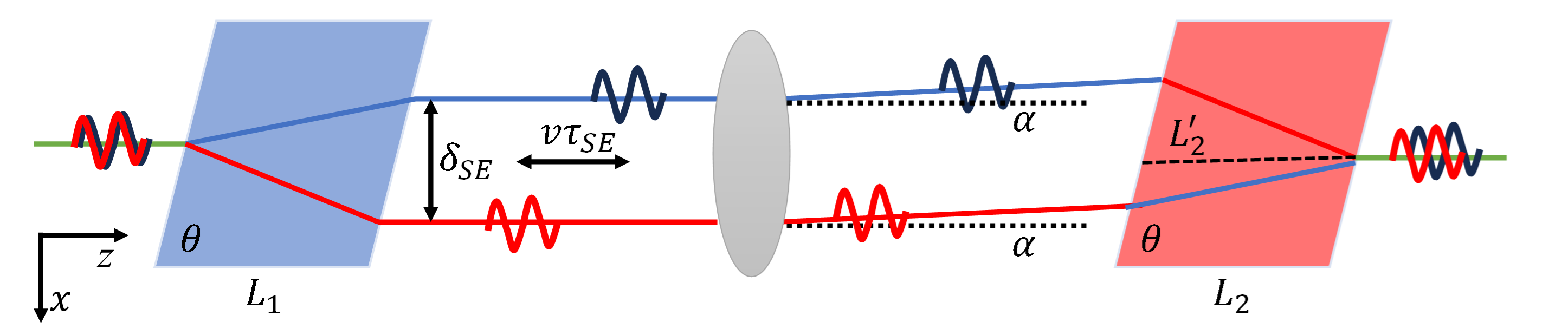}
	\caption{Schematic representation of Spin-Echo Small Angle Neutron Scattering. Two inclined magnetic field regions of opposite polarization (represented in light blue and red) are used to induce a transverse separation between the two neutron spin states, also indicated in blue and red. The SESANS configuration shown here also induces a longitudinal separation between the wavepackets indicated by $v\tau_{SE}$. The sample indicated in gray causes the beam to be scattered by a small angle $\alpha$. As a result the path length through the second field region is changed $L_2'$ (as opposed to the non-scattered path $L_2$), resulting in a net phase shift between the two spin states upon recombination.}\label{SESANS}
\end{figure*}
It can be seen that for short wavelengths the chopper systems perform with equal relative efficiency, however at longer wavelengths the spin chopper has a higher transmission compared to the mechanical chopper. This is due to the wavelength dependent transmission of the mechanical Fermi chopper \cite{Fermi1947} which we may approximate at long wavelengths by 1 minus the ratio between the channel width seen by a neutron of wavelength $\lambda$ and the channel width when the chopper is at rest: 
\begin{equation}
    T=1-\frac{\omega m \lambda D^2}{4\pi d \hbar}
\end{equation}
with $\omega$ the rotation frequency of the device, $D$ the length and $d$ the width of the channels. The transmission of the spin chopper on the other hand is independent of wavelength when the RF pulse time is much shorter than the flight time of the neutron through the RF coil. In this case, much like the double disk chopper \cite{Well2005}, the wavelength resolution is also wavelength independent, depending only on the distance between the chopper and the detector, $L$ and the length of the RF coil $D$
\begin{equation}
    \frac{\Delta \lambda}{\lambda}=\frac{D}{L}
\end{equation}
In addition the RF pulse frequency may be tuned to optimize the available intensity for a given detector distance.
\subsection{Spin Echo Small Angle Neutron Scattering}
Spin Echo Small Angle Neutron Scattering (SESANS) is an interferometric spin echo technique first designed to resolve small angle scattering smaller than the beam divergence \cite{Pynn1978,Rekveldt1996}. Later the interferometric applications were further developed and applied to a variety of questions ranging from quantum contextuality \cite{Shen2020,Kuhn2021} to (exotic \cite{Parnell2020}) gravity \cite{Haan2014} and to probe properities of the neutron itself, such as the intrinsic coherence of a single neutron \cite{McKay2024}. Regular Spin Echo, uses two oppositely magnetized field regions, such that all precession induced in the first field region is reversed in the second, if the neutron velocity does not change in the instrument, ie.
\begin{equation}
    \gamma B_1 \frac{L_1}{v_1}=-\gamma B_2 \frac{L_2}{v_2}
\end{equation}
\begin{figure*}
	\includegraphics[width=18cm]{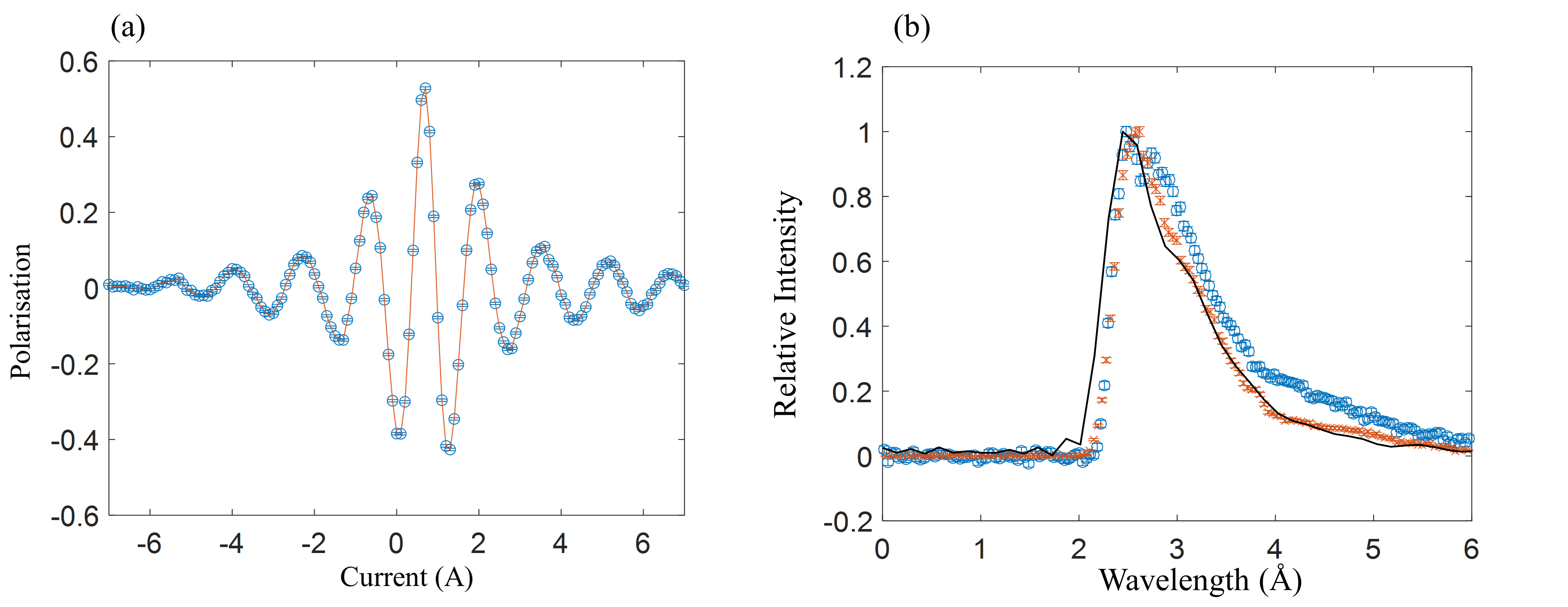}
	\caption{ (a) Polarisation of the spin echo group measured against current in the precession coil. The individual data points with errorbars are shown in blue and in interpolation of the data is shown in red. (b) The Fourier transform of the spin echo group is shown in black, while the spectra measured using both chopping techniques also shown in figure \ref{TOF} is superposed. Note that no sample was inserted in the instrument for these measurements.}\label{SpinEchoGroup}
\end{figure*}
with $B$ the magnetic field strength, $L$, the length of the precession region and $v$ the neutron velocity. The index refers to the first or second precession region.
It follows that such a spin echo instrument is sensitive to small changes of the neutrons kinetic energy. In SESANS the field regions are inclined (see figure \ref{SESANS}), so that the instrument becomes sensitive to small changes in the neutrons transverse wavevector component, $q$ (small angle scattering). If the velocity of the neutron is unchanged the net phase accumulated by the neutron after a small angle scattering event is simply
\begin{equation}
    \chi=\gamma B_2 \frac{\Delta L_2}{v}
\end{equation}
where $\Delta L_2=L_2'-L_2$ is the difference between the scattered and non-scattered neutron flight path length in the second arm, given by
\begin{equation}
    \Delta L_2=L_2\cot(\theta_0) \alpha=L_2\cot(\theta_0)\frac{q}{k}
\end{equation}
with $\alpha$ the scattering angle. It follows that the phase is given by
\begin{equation}
    \chi=\frac{\gamma m B_2 L_2 \cot(\theta_0) \lambda^2}{4\pi^2\hbar}q=\delta_{SE} q
\end{equation}
with the spin-echo length $\delta_{SE}$ equal to the induced transverse separation between the up and down spin state. The polarisation is therefore a cosine transform of the scattering function $S(q)$
\begin{equation}
    P=\int dq \cos(\delta_{SE} q) S(q) = G(\delta_{SE})
\end{equation}
with $G(\delta_{SE})$ the real space correlation function. Since $\delta_{SE}$is proportional to $\lambda$, a white beam SESANS, such as CANISIUS, averages over many different correlation lengths.
\begin{equation}
    P=\frac{\int d\lambda G(\delta_{SE}) I(\lambda)}{\int d\lambda I(\lambda)}
\end{equation}
with $I(\lambda)$ the intensity distribution of the sampled beam. To extract the wavelength information an additional precession coil (see nr. 5 in figure \ref{Setup}), can be used to measure the spin echo group. This precession coil adds a phase which is field and wavelength dependent
\begin{equation}
    \varphi=\frac{\gamma B m \lambda d}{2\pi \hbar}=\kappa\lambda
\end{equation}
with $d$ the length of the precession coil. Hence by adding the precession coil the polarisation becomes the cosine transform of the correlation function times the intensity distribution
\begin{equation}
    P(\kappa)=\frac{\int d\lambda G(\delta_{SE}) I(\lambda) \cos(\kappa\lambda)}{\int d\lambda I(\lambda)}
\end{equation}
It follows that the cosine transform of the polarisation of the spin echo group, $P(\kappa)$ yields the correlation function weighted by the normalized intensity distribution. The correlation function of a specific sample can be extracted by normalizing the cosine transform of the spin echo group with sample, $P_1(\kappa)$ by a cosine transform of the spin echo group without the sample, $P_0(\kappa)$.
\begin{equation}\label{Equation:Correlation}
    G(\delta_{SE})=\frac{\int d\kappa P_1(\kappa) \cos(\kappa\lambda)}{\int d\kappa P_0(\kappa) \cos(\kappa \lambda)}
\end{equation}
Since the correlation function of a vacuum is constant, the cosine transform of the spin echo group without a sample should simply be equal to the normalized spectrum. In figure \ref{SpinEchoGroup} one can see the spin echo group measured by CANISIUS and its Fourier transform superposed on the normalized ToF spectra. \par
\begin{figure}
	\includegraphics[width=9cm]{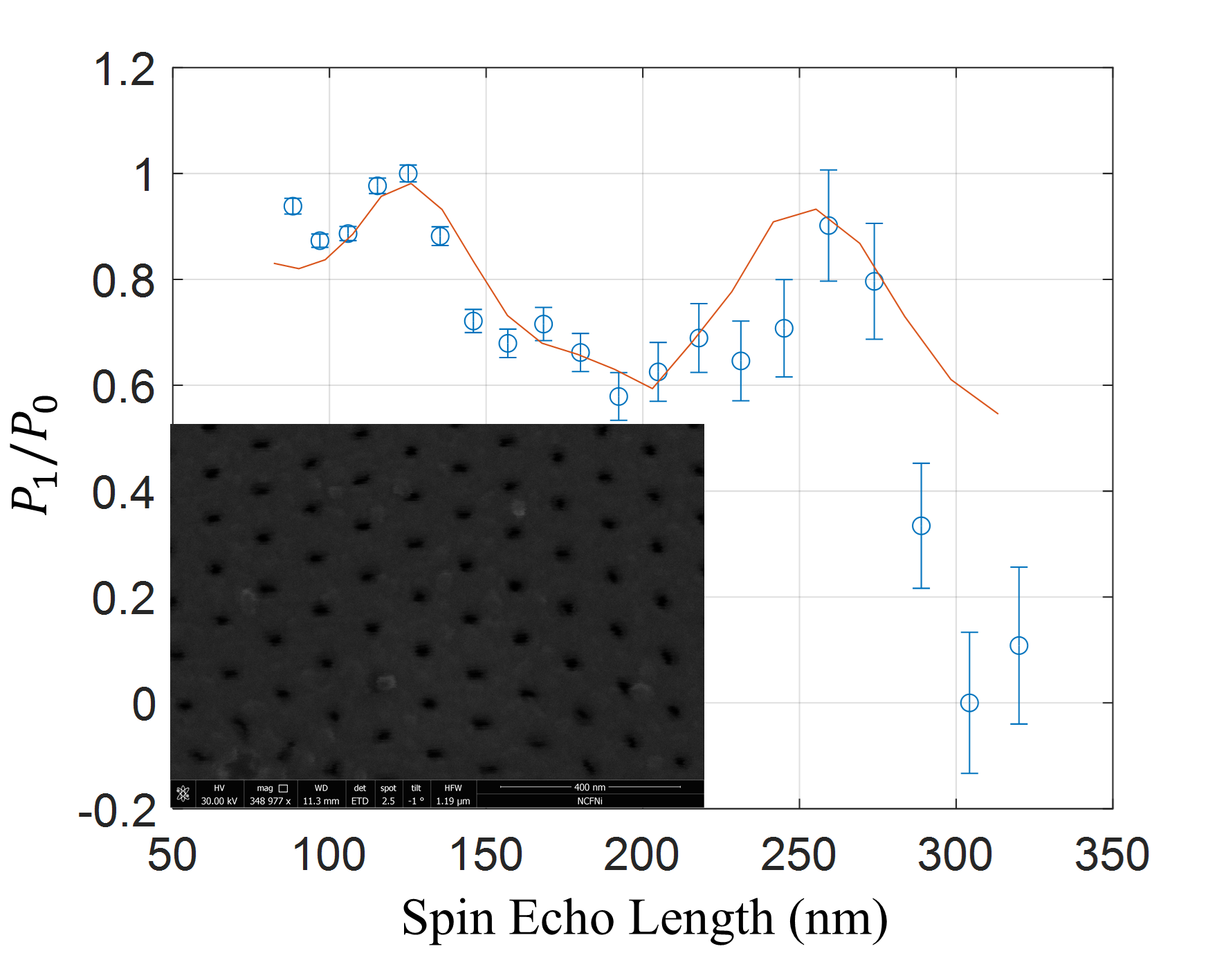}
	\caption{Ratio of Fourier transformed sample in to sample out polarization measured using the white beam method on CANISIUS (blue) superposed on a fit function obtained using the phase object approximation. The inset shows an SEM image of the nanoporous alumina, carried out by USTEM at the TU Wien.}\label{CorrelationFunction}
\end{figure}

 
The spin echo length of CANISIUS was calibrated using a nanoporous alumina membrane from smartmembranes \cite{Smartmembranes}, similar to the method described in \cite{Funama2024}. The pores used have a diameter of $40 \mathrm{nm}$, with a pitch of $125\mathrm{nm}$ and a thickness of $50$ microns. A SEM image of the sample is shown in the inset of figure \ref{CorrelationFunction}. Since the pitch of the pores are precisely known we can use the measured SESANS correlation function to calibrate the spin echo length of CANISIUS. Note that the sample and sample holder confine the beam size to  7 x 10 mm$^2$. Figure \ref{CorrelationFunction} shows the correlation function 
\begin{figure*}
	\includegraphics[width=18cm]{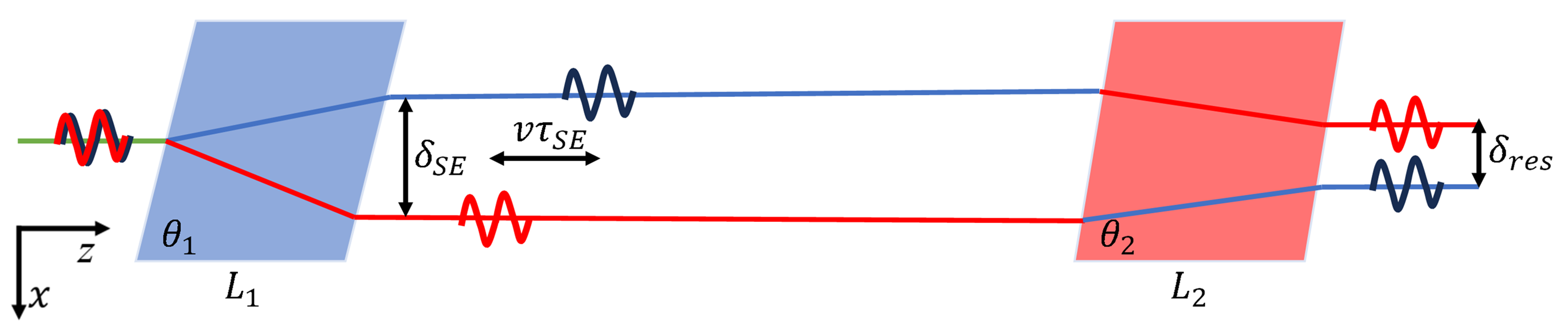}
	\caption{Schematic representation of the coherent averaging mode of the CANISIUS instrument. This mode is almost equivalent to SESANS, the only difference being the different angles between the two field regions. This leads to an incomplete focusing of the two spin states, which is indicated by the residual distance $\delta_{res}$ between the states at the end of the instrument.}\label{CoherentAveraging}
\end{figure*}
measured using CANISIUS, as defined in equation \ref{Equation:Correlation}, superposed on a predicted correlation function based on the phase object approximation \cite{Haan2007}. The phase object approximation is used to calculate the SANS pattern produced by our sample, which in turn is Fourier transformed to obtain a van Hove correlation function $g(\delta_{SE})$. Finally we can convert this van Hove correlation function into a polarization ratio as described by equation \ref{Equation:Correlation}, using the relation $G(\delta_{SE})=e^{\Sigma t (1-g(\delta_{SE})}$, with $t$ the sample thickness and $\Sigma$ the macroscopic scattering cross section. By making use of the phase object approximation, in particular the SANS pattern, we can easily model the effect of instrument acceptance on the observed correlation function. The spin echo length axis of the CANISIUS measurement shown in in figure \ref{CorrelationFunction} has been scaled by a factor of 0.65 to overlap with the phase object approximation.
Overlap between the white beam CANISIUS measurement and the estimate of the correlation function produced using phase object approximation demonstrates the feasibility of the white beam method detailed in this work. Our measurement shows that the spin echo length is shorter by a factor of 0.65 compared to the theoretical estimate based on the applied fields and distances. This discrepancy is most easily explained by the large stray fields between the RF flippers and the V-coils, which reduces the total amount of zero field precession.
\begin{figure*}
	{\includegraphics[width=18cm]{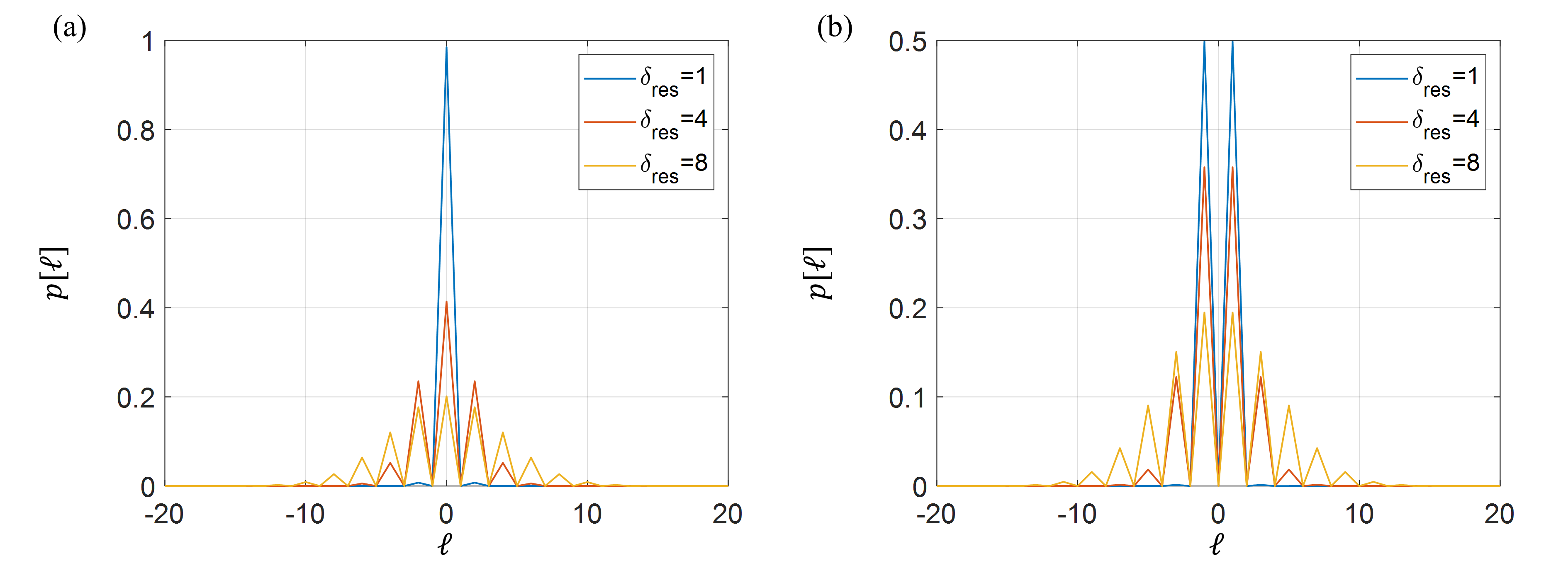}}\caption{OAM distribution function eq. \ref{PWTOAM2}, in the special case where $\sigma=1$ (a) $\beta=0$ and (b) $\beta=\pi$ for various residual spin echo lengths $\delta_{res}$. The OAM is measured around the propagation axis z. For both $\beta$ the OAM amplitude is non-zero only for every other mode. In (a) only even modes contribute, while in (b) only odd modes make up the wavefunction.}\label{LongCase}
\end{figure*}
\subsection{Coherent Averaging}
Constructing structured wavefunctions by coherent averaging is the main purpose of the CANISIUS instrument. This is done primarily to produce neutron wavefunctions that have non zero Orbital Angular Momentum (OAM) and investigate the properties of said states. Put simply coherent averaging is an interferometric technique that takes an input wavefunction with simple structure (i.e. Gaussian) and splits it multiple identical copies, called partial wavefunctions. These partial wavefunctions can then be independently translated and phase shifted in real or reciprocal space to produce a composite wavefunction (i.e. the coherent sum of all partial wavefunctions) that exhibits the desired structure. This technique is particularly well suited for producing neutron OAM, as explored in \cite{Sarenac2018,Sarenac2018b,Sarenac2019,Geerits2023}. Coherent averaging could be used to produce a circular array of partial wavefunctions which act as coherent sources. If one can control the phase of each source, it becomes possible to produce arbitrary OAM values as described in \cite{wang2023topology}. We have demonstrated the method previously in perfect crystal neutron interferometry \cite{Geerits2023}. \par
\begin{figure*}
	\centering
	{\includegraphics[width=18cm]{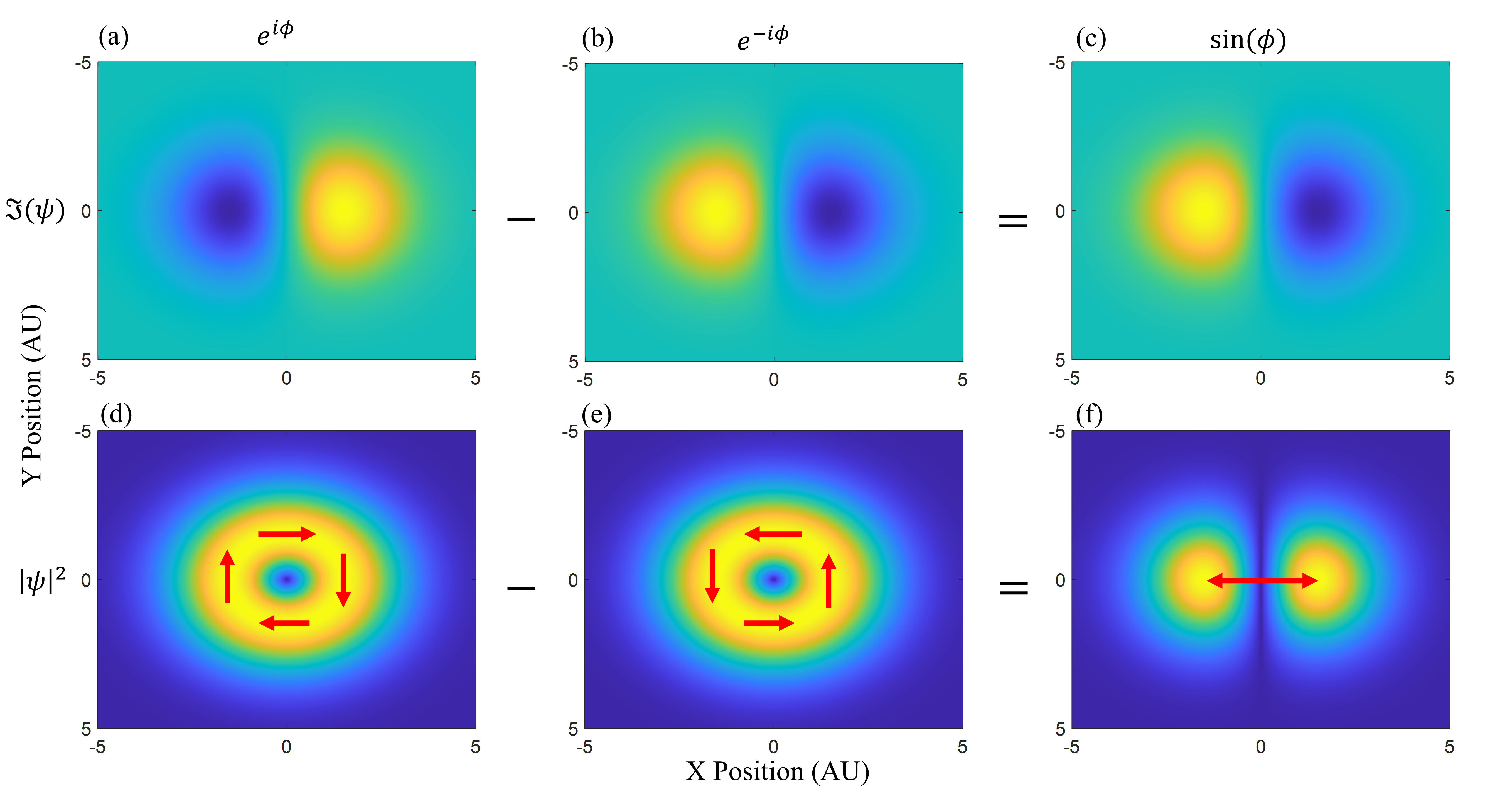}}\caption{Illustration of the construction of a linearly polarized OAM state. A clockwise rotating state (a)/(d) has a counter clockwise rotating state (b)/(e) subtracted from it, producing a linearly polarized state (c)/(f). The top insets (a)-(c), show the imaginary parts of the respective transverse wavefunction, while the bottom insets (d)-(f) show the absolute values, which is directly related to the expected intensity pattern.}\label{LinOAM}
\end{figure*}

 
Since SESANS is also a type of neutron interferometer it can be used to produce structured wavefunctions by coherent averaging. Here the structured wave is produced by running the instrument in a new mode, where the first and second arm have different poleshoe angles. As a result the two spin states retain some transverse separation after passing through the spin echo interferometer. The concept is illustrated in figure \ref{CoherentAveraging}. This residual transverse separation is given by
\begin{equation}
    \delta_{res}=\frac{\gamma m B L\lambda^2}{4\pi^2\hbar}(\cot(\theta_1)-\cot(\theta_2))
\end{equation}
where we have assumed that $B=B_1=B_2$ and $L=L_1=L_2$. So for a Gaussian input wavefunction
\begin{equation}
    \psi_0 = A e^{-x^2/\sigma^2}e^{ik_y y}
\end{equation}
with $\sigma$ the transverse coherence length the transvers part of the output wavefunction after spin selection would be equal to
\begin{equation}\label{GaussianPW2}
	\psi_{1}=\frac{A}{\sqrt{2}} (e^{\mathrm{i}\beta/2}e^{-\frac{(x-\delta_{res}/2)^2-y^2}{\sigma^2}}+e^{-\mathrm{i}\beta/2}e^{-\frac{(x+\delta_{res}/2)^2-y^2}{\sigma^2}})
\end{equation}
with $\beta$ a longitudinal phase between the two partial wavefunctions. $\beta$ can be precisely tuned by changing the phase of one of the RF coils $\beta=2\Delta\phi$. We can analyze the OAM or mode distribution by introducing the azimuthal Fourier transform
\begin{equation}\label{AFT}
    \psi^\ell(r,z) = \frac{1}{2\pi} \int_0^{2\pi}\psi(r,\phi,z) e^{-\mathrm{i}\ell\phi} \mathrm{d}\phi
\end{equation}
and its inverse
\begin{equation}
    \psi(r,\phi,z) = \sum_\ell \psi^\ell(r,z) e^{\mathrm{i}\ell\phi}
\end{equation}
When we apply the azimuthal Fourier transform to our composite wavefunction we find
\begin{equation}\label{Vortex_PW2}
	\psi_{1}^\ell(r)=\sqrt{2}Ae^{-\frac{r^2+\delta_{res}^2/4}{\sigma^2}} J_\ell (\mathrm{i}\frac{\delta_{res}}{\sigma^2}r) \cos(\frac{\pi \ell+\beta}{2})
\end{equation}
Since this mode amplitude is complex and spatially dependent it can be useful to introduce a mode distribution function defined as
\begin{equation}
    p[\ell] = 2\pi \int r|\psi^\ell(r)|^2 \mathrm{d}r
\end{equation}
which is related to the OAM expectation value through
\begin{equation}
    <\hat{L}_z> = \hbar \sum_\ell \ell p[\ell]
\end{equation}
When we calculate the mode distribution function of \ref{Vortex_PW2} we find
\begin{equation}\label{PWTOAM2}
	p[\ell]= \frac{A^2\sigma^2}{2}e^{-\frac{\delta_{res}^2}{4\sigma^2}} I_\ell(\frac{\delta_{res}^2}{4\sigma^2})|\cos|^2(\frac{\pi \ell+\beta}{2})
\end{equation}
We may now observe that for $\beta=0$, $p[\ell]$ is only non zero for even $\ell$, that is to say only the even OAM modes contribute to the output wavefunction, while for $\beta=\pi$ $p[\ell]$ is only non zero for odd $\ell$ and hence only odd OAM modes contribut to the output wavefunction. By tuning the ratio $\delta_{res}/\sigma$ we can determine how many modes contribute. For small splitting only few modes contribute whereas for large splitting many modes shape the wavefunction. The OAM distribution function along with these properties are shown in figure \ref{LongCase}. Interestingly we find that for $\beta=\pi$ and small $\delta_{res}$ with respect to the coherence length the output wavefunction can almost entirely be described by the $\ell=-1$ and $\ell=1$ modes, i.e. a right and left twisting mode, somewhat analogous to linear polarization. The concept of a "linear" OAM state is illustrated in figure \ref{LinOAM}. 
Here a left rotating state is subtracted from a right rotating state to produce a linearly polarized (sine oscillation) state. We have now shown that such a linearly polarized state can be approximated by a superposition of two Gaussian wavepackets with opposite amplitudes on opposite sides of the cylinder axis (eq. \ref{GaussianPW2}). If $\delta_{res}$ is smaller than or on the order of $\sigma$, we find that figure \ref{LongCase} and equation \ref{PWTOAM2} demonstrate that this very closely approximates a linearly polarized OAM state with $|\ell|=1$.
While this state has an average OAM of zero just like a pure $\ell=0$ mode it is orthogonal to a pure $\ell=0$ state as nicely illustrated by looking at the OAM distribution function. In addition, even though the average OAM of each wavefunction produced by CANISIUS by coherent averaging is zero, the superpositions of $\ell=\pm 1$ or $\pm 2$ are qualitatively different than the pure $\ell=0$ mode, which are expected to scatter differently from nuclear targets \cite{Afanasev2019,Afanasev2021}. In addition the absorption characteristics of these states are expected to be different \cite{Jach2022}. And as described in a recent paper these states may be used to explore the quantum Sagnac effect \cite{Geerits2024}, where these superposition states may produce intensity beating proportional to the mode number and the rotation frequency of the observer.
\section{Conclusion}
We have presented the new Austrian spin echo interferometer of the Atominstitut, CANISIUS, and illustrated its capabilities, both as a ToF and a broad-band white beam instrument. In particular we demonstrated the feasibility of polychromatic SESANS, which allows us to use the full potential of a continuous neutron source. This is especially important at weaker sources such as the 250 kW Triga reactor at the Atominstitut. Finally the coherent averaging capabilities of the instrument were illustrated. It was shown that CANISIUS is capable of producing neutrons that carry non zero OAM, in particular states that exist in a superpostion of the $\ell=\pm 1$ modes. We expect this will be useful for exploring OAM dependent scattering and absorption cross sections, in addition to exploring fundamental aspects of quantum mechanics from the quantum Sagnac effect to quantum information and contextuality.
\section{Acknowledgments}
N. G. and S. S. acknowledge funding from the Austrian
science fund (FWF), Project No. P34239, in addition
N. G. is supported by the US Department
of Energy (DOE) grant DE-SC0023695. The thermal white beam facility was designed, constructed and funded by the Federal Ministry of Education, Science and Research of Austria. Finally the authors acknowledge and thank Wim Bouwman for useful discussion and resources regarding white beam SESANS.
\section{References}
\bibliographystyle{unsrt}
\bibliography{SpinOrbitBibliography}
\end{document}